\documentclass[twocolumn]{emulateapj}

\usepackage{latexsym,amssymb}
\usepackage{amsmath}
\usepackage{natbib}
\usepackage{rotating}

\usepackage[colorlinks=true,linkcolor=blue,citecolor=blue]{hyperref}
\usepackage{tikz}
\def\checkmark{\tikz\fill[scale=0.4](0,.35) -- (.25,0) -- (1,.7) -- (.25,.15) -- cycle;}

\newcommand{\mysim}{\mathord{\sim}}

\newcommand{\myapprox}{\mathord{\approx}}

\newcommand{\mwd}{M_{\textrm{WD}}}
\newcommand{\mhe}{M_{\textrm{He}}}
\newcommand{\mcross}{M_{\textrm{He,cr}}}
\newcommand{\msteady}{M_{\textrm{He,st}}}
\newcommand{\mevol}{M_{\textrm{He,ev}}}

\slugcomment{Draft \today}
\shorttitle{Collisions}
\shortauthors{Holcomb et al.}

\begin{document}

\title{Can helium envelopes change the outcome of direct white dwarf collisions?}
\author{Cole Holcomb\altaffilmark{1}, Doron Kushnir \altaffilmark{2,3}}

\altaffiltext{1}{Department of Astrophysical Sciences, Princeton University, Princeton, NJ 08540, USA}
\altaffiltext{2}{Institute for Advanced Study, Princeton, NJ 08540, USA}
\altaffiltext{3}{John N.\ Bahcall Fellow}

\begin{abstract}

Collisions of white dwarfs (WDs) have recently been invoked as a possible mechanism for type Ia supernovae (SNIa). A pivotal feature for the viability of WD collisions as SNIa progenitors is that a significant fraction of the mass is highly compressed to the densities required for efficient $^{56}$Ni production before the ignition of the detonation wave. Previous studies have predominantly employed model WDs composed entirely of carbon-oxygen (CO), whereas WDs are expected to have a non-negligible helium envelope. Given that helium is more susceptible to explosive burning than CO under the conditions characteristic of WD collision, a legitimate concern is whether or not early time He detonation ignition can translate to early time CO detonation, thereby drastically reducing $^{56}$Ni synthesis. We investigate the role of He in determining the fate of WD collisions by performing a series of two-dimensional hydrodynamics calculations. We find that a necessary condition for non-trivial reduction of the CO ignition time is that the He detonation birthed in the contact region successfully propagates into the unshocked shell. We determine the minimal He shell mass as a function of the total WD mass that upholds this condition. Although we utilize a simplified reaction network similar to those used in previous studies, our findings are in good agreement with detailed investigations concerning the impact of network size on He shell detonations. This allows us to extend our results to the case with more realistic burning physics. Based on the comparison of these findings against evolutionary calculations of WD compositions, we conclude that most, if not all, WD collisions will not be drastically impacted by their intrinsic He components. 

\end{abstract}

  
\section{Introduction}\label{sec:int} 

Type Ia supernovae (SNIa) are well-known cosmological ``standardizable candles'' thanks to a tight empirical correlation (the Phillips relation; \citealt{1993ApJ...413L.105P}). It is understood that SNIa are powered by the decay of $^{56}$Ni \citep[][]{1969ApJ...157..623C} produced from the explosion of White Dwarfs (WDs) composed predominantly of carbon and oxygen (CO), but there is no consensus regarding the explosion mechanism. The two canonical scenarios, single-degenerate accretion (WD accretion exceeding the Chandrasekhar limit) and double-degenerate mergers (merger of two close WDs that spiral in due to gravitational radiation), have many theoretical and observational challenges \citep{2000ARA&A..38..191H,2014ARA&A..52..107M}. A serious concern for both scenarios is that a successful ignition of an explosive detonation has never been convincingly demonstrated \citep[including the recent claims of a self-consistent ignition in a double-degenerate merger;][see section~\ref{sec:num} for details]{2015ApJ...800L...7K}.

Although collisions of WDs were believed to have rates which are orders of magnitude smaller than the rate of SNIa, they motivated three-dimensional (3D) hydrodynamic simulations of such collisions and of the resulting thermonuclear explosion \citep{1989ApJ...342..986B,2009MNRAS.399L.156R,2009ApJ...705L.128R,2009A&A...500.1193L,2010ApJ...724..111R,2012ApJ...759...39H,2013MNRAS.434.2539A}.While the amount of $^{56}$Ni synthesized in most of these simulations was non-negligible, the results were contradictory, with inconsistent amounts of $^{56}$Ni and different ignition sites of a detonation wave for the same initial conditions. These discrepancies were resolved by \citet{2013ApJ...778L..37K}, where high-resolution two-dimensional (2D) simulations with a fully resolved ignition process were employed. The nuclear detonations in these collisions are due to a well understood shock ignition that is devoid of the commonly introduced free parameters such as the deflagration velocity or transition to detonation criteria \citep[e.g., in the single-degenerate and double-degenerate scenarios; see ][]{2000ARA&A..38..191H}.

\cite{2012arXiv1211.4584K} demonstrated that the rate of direct collisions in common field triple systems may approach the SNIa rate. \cite{2011ApJ...741...82T} had previously argued that the secular Lidov-Kozai mechanism \citep{1962AJ.....67..591K,1962P&SS....9..719L} in triples might play an important role in WD-WD mergers via gravitational radiation to produce SNIa. However, the non-secular corrections to the Lidov-Kozai mechanism obtained by \cite{2012arXiv1211.4584K} raised the possibility that the majority of SNIa result from collisions.
Supporting evidence was provided in \cite{2013ApJ...778L..37K}, in which numerical simulations reproduced several robust observational features of SNIa. In particular, it was established that the full range of $^{56}$Ni necessary for all SNIa across the Phillips relation can be obtained by collisions of typical WDs. Further evidence was recently discovered by \cite{2015MNRAS.454L..61D}  in the form of doubly-peaked line profiles in high-quality nebular-phase spectra, which suggest that SNIa with intrinsic bi-modality are common. They observe such bi-modality in a 3D $0.64 M_{\odot}$-$0.64 M_{\odot}$ WD collision simulation as a result of detonation in both WDs.

A crucial property for the viability of WD collisions as progenitors of SNIa is that a significant fraction of the mass is highly compressed to the densities required for efficient $^{56}$Ni production \emph{before} the ignition of the detonation wave. Otherwise only massive ($>0.9\,M_{\odot}$) CO WDs would produce sufficent quantities of $^{56}$Ni (as required by all other progenitor models). Evolutionary calculations predict that CO WDs retain helium in the outermost layers \citep{2005A&A...435..631A,2006MNRAS.371..263L,2010ApJ...717..183R}, see Figure \ref{fig:param}. A legitimate concern, then, is whether or not burning of the helium envelope (which is possible at lower temperatures than required for CO burning) can lead to a CO ignition at early times, before a significant fraction of the mass is highly compressed. The predictions for the helium envelope mass, $\mevol$, are between $\myapprox10^{-3}\,M_{\odot}$ and $\myapprox2.5\times 10^{-2}\,M_{\odot}$ for $0.9\,M_{\odot}$ and $0.5\,M_{\odot}$ WDs, respectively \citep[see, however, significantly less massive helium shells from asteroseismic analysis of helium-atmosphere (DB) white dwarfs;][]{2014ApJ...794...39B}. 

The impact of helium shells was recently studied by \citet{Papish:2015uy}, who performed 2D head-on collision calculations. They scanned the relevant parameter space with 8 collisions\footnote{We ignore the equal mass collisions with different helium masses of \citet{Papish:2015uy}. The helium shell masses must be equal because the WDs are at the same age and at the same metallicity.}, and concluded that (see Figure \ref{fig:param})
\begin{itemize}
\item $\mhe\gtrsim 0.1\,M_{\odot}$ is required for $\mwd=0.7\,M_{\odot}$ and $\mwd=0.8\,M_{\odot}$ in order for the detonation to propagate in the helium shell,
\item $\mhe\gtrsim 0.2\,M_{\odot}$ is required for $\mwd=0.8\,M_{\odot}$ in order to obtain CO ignition at early times.
\end{itemize}
These results suggest that the required helium shell masses to obtain CO ignition at early times are much higher than $\mevol$ for the relevant WD masses.\footnote{\citet{Papish:2015uy} suggest that mass transfer can significantly increase $\mhe$ beyond $\mevol$ prior to collision in quadruple systems. Since collisions require wide binaries, mass transfer is unlikely prior to collision, so we ignore this possibility.} However, the scan of the parameter space was quite sparse, resulting in coarse estimates for the minimal helium mass that can alter the ensuing CO ignition, especially for low mass WDs. Perhaps most importantly, they used the approx19 reaction network \citep{1978ApJ...225.1021W} in their simulations, which is known to be a poor representation of helium burning at high temperatures because it does not include the proton mediated $\alpha$-capture reaction $^{12}\textrm{C}(p,\gamma)^{13}\textrm{N}(\alpha,p)^{16}\textrm{O}$ \citep{2006ApJ...639.1018W,2009ApJ...699.1365S,2011ApJ...734...38W,2014ApJ...797...46S}. It is not clear how their results depend on this approximation; for example, we show that their failure to obtain a He shell detonation for $\mwd=0.724\,M_{\odot},\,\mhe=2.4\times 10^{-2}\,M_{\odot}$ case may be due to the limitations of the nuclear network.

We perform a careful study of the role of helium shells in equal-mass head-on collisions (zero impact parameter), by using 2D numerical simulations. Since capturing the dynamics of such collisions requires multi-dimensional simulations, we work with a small reaction network of 13 isotopes (similar to the approx19 reaction network, see below). We demonstrate that a necessary condition for early CO ignition is a stable detonation propagation in the WD helium shell. Establishing this condition allows us to use previous detailed studies of detonation waves in helium shells \citep{2012ApJ...755....4T,2013ApJ...776...97M,2014ApJ...797...46S} to derive a lower limit for the the required helium mass for early CO ignition, which we then extend to the case of large reaction networks. We argue that this lower limit is applicable to all collisions, including unequal mass collisions and those with non-zero impact parameter. We obtain this lower limit for a wide range of WD masses, and we conclude that early CO ignition due to He burning is unlikely in real collisions given the predicted helium shell masses from evolutionary calculations, except possibly for a very small fraction of SNIa at the faint end. We argue that, for $\mhe$ below this lower limit, the $^{56}$Ni distribution in the ejecta should not diverge significantly from the pure CO case. We do not study the synthesis of intermediate mass elements in the burnt helium shell because the yields resulting from the small reaction network utilized here are subject to large uncertainties (previous works are similarly uncertain).

In section \ref{sec:num} we describe the numerical methods used throughout this study. In section \ref{sec:results} we investigate the dynamics of collisions with helium shells and establish a necessary condition for early CO ignition by measuring the lower limit on the requisite helium mass. We conclude in section \ref{sec:sum}.

\section{Numerical Methods \& Setup}\label{sec:num}

We calculate head-on collisions of equal mass WDs with the FLASH\footnote{Version 4.2.2} \citep{2000ApJS..131..273F} hydrodynamics code. The fluid equations are evolved by the directionally split hydrodynamics solver and are closed with the tabular Helmholtz equation of state \citep{2000ApJS..126..501T}. Compositions are updated via a 13 isotope $\alpha$-chain reaction network (similar to the approx13 network supplied with FLASH with slightly updated rates for specific reactions, especially fixing a typo for the reaction $^{28}\textrm{Si}(\alpha,\gamma)^{32}\textrm{S}$, which reduced the reaction rate by a factor $\myapprox4$. Note that the approx19 network does not significantly change the results of the approx13 network for CO and helium burning). The gravitational interaction is calculated by the ``new multipole solver" \citep{2013ApJ...778..181C}, with the multipole expansion out to $l_{\rm max} = 16$. We find that our results are converged when employing adaptive mesh refinement with $\myapprox4$ km resolution (i.e. the minimal allowed cell size within the most resolved regions), see convergence study in section~\ref{sec:results}. 

False numerical ignition may occur if the burning time, $t_{\textrm{burn}}=\dot{Q}/\varepsilon$ (where $\dot{Q}$ is the energy injection rate from burning and $\varepsilon$ is the internal energy), in a cell becomes shorter than the sound crossing time, $t_{\textrm{sound}}=\Delta x/c_{s}$ (where $\Delta x$ is the length scale of the cell and $c_{s}$ is the sound speed). To evade this pitfall, we include a burning limiter that forces the burning time in any cell to be longer than the cell's sound crossing time by suppressing all burning rates with a constant factor whenever $t_{\textrm{sound}}>f t_{\textrm{burn}}$ with $f=0.1$ \citep[see][for a detailed description]{2013ApJ...778L..37K}. In order to illustrate the necessity of such a limiter we analyze the recent claims of a self-consistent ignition in a double-degenerate merger \citep{2015ApJ...800L...7K}. They used the FLASH code without implementing the limiter, and they obtained a CO ignition at a density of $\myapprox6.7\times10^{6}\,\textrm{g}\,\textrm{cm}^{-3}$ and at a temperature of $\myapprox3.2\times10^{9}\,\textrm{K}$. Under these conditions the burning time is $t_{\textrm{burn}}\approx1.4\times10^{-5}\,\textrm{s}$ (for $50\%$ carbon, $50\%$ oxygen, by mass) while the sound crossing time for their highest resolution ($68.3\,\textrm{km}$) is $1.3\times10^{-2}\,\textrm{s}$. Since the burning time is shorter by three orders of magnitude from the sound crossing time the obtained ignition is not physical, and the claim of a self-consistent ignition is not rigorous.

To isolate the behavior of the helium component we utilize a simple model. Isothermal white dwarf profiles\footnote{http://cococubed.asu.edu/code\_pages/adiabatic\_white\_dwarf.shtml} are constructed with temperature $T=10^7$ K. Two regions are then defined: a $50\%$ (by mass) carbon, $50\%$ oxygen core, and a pure helium envelope. The radius of the composition boundary is altered to achieve the desired He envelope mass. Head-on (zero impact parameter) collisions allow the use of cylindrical geometry $(r,z)$. The WDs are intialized in contact with free-fall velocities. The ambient medium consists of helium gas at density $\rho_{\rm amb} = 10^{-2}$ g cm$^{-3}$ and $T_{\rm amb}=10^7$ K. The domain boundaries are $r=[0,L]$ and $z=[-L,L]$, where $L=2^{17}\,\textrm{km}$ $\myapprox$ $1.31\times 10^{5}\, \textrm{km}$.

In order to broadly probe the parameter space at hand, we choose the WD mass pairs (in units of $M_{\odot}$) 0.5-0.5, 0.64-0.64, and 0.8-0.8; see Tables \ref{table:toy}, \ref{table:toy2}, and \ref{table:toy3} for a summary of the models and their main results. The results from the pure CO models are consistent with the previous results of \cite{2013ApJ...778L..37K}, and provide a means of understanding the role of He layers in WD collisions by comparison.

\section{Results}\label{sec:results}

In this section we investigate the dynamics of collisions with helium shells and establish a necessary condition for early CO ignition. We begin by examining the $0.64-0.64$ case in Section~\ref{sec:0.64-0.64}. The lower limit for the required helium mass for early CO ignition is presented in Section~\ref{sec:calib}.

\subsection{The $0.64-0.64$ case}\label{sec:0.64-0.64}

The dynamical evolution of the collisions is shown in Figure \ref{fig:shelldet} for four different values of $\mhe$: $0$, $4\times10^{-2}\,M_{\odot}$, $8\times10^{-2}\,M_{\odot}$ and $0.16\,M_{\odot}$, with $M_{\rm WD} = 0.64\,M_{\odot}$. These representative collisions demonstrate different behavioral regimes, depending on the helium mass. Two strong shocks (the leading shocks hereafter) initially propagate from the contact surface and move toward the center of each star at a velocity that is a small fraction of the velocity of the approaching stars (the fact that the stars are identical implies a mirror symmetry $\pm z$ allowing us to focus on one of the stars in Figure \ref{fig:shelldet}). The shocked region near the contact surface has an approximately planar symmetry and a nearly uniform pressure \citep{2014ApJ...785..124K}. The temperature near the surface of contact is too low for appreciable nuclear burning to take place at early times for the pure CO case (panel (a1)), however significant burning of the shocked helium shell is obtained after $\mysim0.6\,\textrm{s}$ following the collision in the non-zero $M_{\rm He}$ cases (panels (b1-d1)). The induction time for the helium burning at this stage is calculated accurately with our small network because the temperatures are below $10^{9}\,\textrm{K}$, where the $^{12}\textrm{C}(p,\gamma)^{13}\textrm{N}(\alpha,p)^{16}\textrm{O}$ reaction is not important. The helium burning increases the pressure in the helium shell and accelerates the leading shock (compare the position of the shock in panels (a1-d1)). The burning also leads to an ignition of a detonation wave for massive enough ($\geq$$8\times 10^{-3}\,M_{\odot}$) helium shells. We emphasize that the ignition is completely resolved in our simulations, and is not put in by hand as in other models. The detonation wave propagates along the $r$-direction inside the shocked helium shell (panels (b1-d1)). Beginning from this phase the helium burning is no longer calculated accurately within the small reaction network \citep{2014ApJ...797...46S}.

The acceleration of the leading shock due to the helium burning is a small effect compared to the acceleration caused by the gravitational field of each star, and therefore the leading shocks continue to accelerate roughly as in the pure CO case until CO ignition is obtained (at $\myapprox2.56\,\textrm{s}$ after contact for the pure CO case, panel (a2); ignition is defined as a formation of a shock due to thermonuclear burning). However, the dynamics can change appreciably if the detonation wave in the shocked helium shell can cross into and propagate within the unshocked helium shell. In the case of $\mhe=4\times10^{-2}\,M_{\odot}$ the detonation wave does not cross into the helium shell (panel (b2)), and the CO ignition is obtained roughly at the same time and location as in the pure CO case (at $\myapprox2.51\,\textrm{s}$ after contact, panel (b2)). For the $\mhe=8\times10^{-2}\,M_{\odot}$ case the detonation wave \emph{does} cross into the unshocked helium shell, propagating to the posterior of the WD. Nevertheless, the CO ignition is obtained roughly at the same time and location as in the pure CO case (at $\myapprox2.43\,\textrm{s}$ after contact, panel (c2)). 

At an even larger helium mass, $\mhe=0.16\,M_{\odot}$, the shock wave launched from the helium shell detonation wave into the CO core converges on the WD interior, leading to a CO ignition before the ignition behind the leading shock (panel (d2), similar to the double-detonation scenario \citep{1990ApJ...354L..53L,2013ApJ...774..137M,2014ApJ...785...61S}). In this particular case the CO ignition is obtained roughly at the same time as in the pure CO case (at $\myapprox2.68\,\textrm{s}$ after contact), however the position is significantly different. The total $^{56}$Ni yield will be similar to the pure CO case because significant fractions of the colliding WDs are allowed to compress to high densities, but we expect large discrepencies in the $^{56}$Ni distribution. For more massive helium shells the detonation wave traverses the helium shell faster (because of the smaller circumference at the composition interface), so that the CO ignition due to the converged shock happens earlier, resulting in drastic reductions to the $^{56}$Ni yield.

From the analysis presented so far, we conclude that a necessary condition for early CO ignition is that the He detonation wave crosses into the unshocked helium shell. In other words, a collision with a given $\mhe$ cannot significantly depart from the pure CO case if it cannot also drive a detonation through the unshocked He shell. We find the minimal helium shell mass that allows this crossing, $\mcross$, to be $(66.5\pm0.5)\times 10^{-3} M_{\odot}$ in the $M_{\rm WD}=0.64\,M_{\odot}$ case, where a successful crossing is declared if a steady detonation wave is propagating in the helium shell (or the entire helium shell is burnt) at the time of CO ignition. For all simulations with $\mhe<\mcross$ the CO ignition was obtained roughly at the same time and location. The same calibrated $\mcross$ was obtained for resolutions of $8\,\textrm{km}$ and $16\,\textrm{km}$. Therefore our results for $\mcross$ (for our small reaction network) are converged to the level of $\myapprox10^{-3}\,M_{\odot}$.  

\subsection{The lower limit on the required helium mass for early CO ignition}\label{sec:calib}

As in Section \ref{sec:0.64-0.64}, we measured $\mcross$ for the $0.5-0.5$ and $0.8-0.8$ cases, and the results are $(93.5\pm0.05)\times 10^{-3} M_{\odot}$ and $(39.5\pm0.5)\times 10^{-3} M_{\odot}$, respectively (Figure \ref{fig:param}). For the $0.5-0.5$ case a similar behavior to the $0.64-0.64$ case is obtained: the CO ignition time becomes slightly later as $\mhe$ approaches $\mcross$. Beyond the transition point to the double detonation-like ignition mechanism, the CO ignition time becomes earlier as $\mhe$ increases. We note that CO ignition due to the converging shock happens for all $\mhe\ge\mcross$ in the $0.5-0.5$ case. The behavior in the the $0.8-0.8$ case is similar to the $0.64-0.64$ case for $\mhe<\mcross$. For $\mhe>\mcross$ the CO ignition is delayed, until for massive enough shells ($\mhe\gtrsim0.18 M_{\odot}$) the helium detonation directly ignites the CO on the symmetry-axis at early times. This is different from the CO ignition due to the converged shock behavior described by \citet{Papish:2015uy} for the $\mwd=0.8\,M_{\odot},\,\mhe=0.2\,M_{\odot}$ case. Nevertheless for $\mhe<\mcross$ the CO ignition was obtained roughly at the same time and location. 

The $\mcross$ that we observe should not be too far from the minimal helium shell mass that allows steady detonation propagation, $\msteady$, which was calibrated in \cite{2013ApJ...776...97M} and \cite{2014ApJ...797...46S} both for the approx13 network, $M^{\alpha}_{\rm He,st}$, and for a large network (in this case, 206 isotopes) with all relevant reactions, $M^{206}_{\rm He,st}$. We find that $\mcross$ is in good agreement with the approx13 determination of $M^{\alpha}_{\rm He,st}$ (Figure \ref{fig:param}), independently confirming the results of their 1D calculations. Therefore the values of $M^{206}_{\rm He,st}$ determined by a large network should provide a good estimate for the influence that the small network approximation had on our results. Figure \ref{fig:param} shows that $M^{206}_{\rm He,st}$ is still significantly above $\mevol$ for the range that it was calculated ($\mwd\ge0.6\,M_{\odot}$). \cite{Papish:2015uy} calculated a collision with $M_{\rm WD} = 0.724\,M_{\odot}$ and $M_{\rm He} = 0.024\,M_{\odot}$ and found that He detonations did not propagate into the unshocked shell. Since these masses are in close proximity to $M^{206}_{\rm He,st}$, this result is rendered uncertain due to their use of an $\alpha$-chain network. Extrapolating $M^{206}_{\rm He,st}$ to $\mwd=0.5\,M_{\odot}$ we can estimate that it is still above $\mevol$ by a few tens of percent. These low mass WDs can only be responsible for a  very small fraction of SNIa at the faint end \citep[see Figure \ref{fig:param}, which shows that the vast majority of CO WDs have $\mwd\ge0.55\,M_{\odot}$][]{2008AJ....135.1225H}.

Although all collision calculations conducted in this study are 2D (zero impact parameter) and employ equal mass WD models, we expect that the same behavior will be obtained in non-zero impact parameter and/or unequal mass collisions. Previous detailed studies of detonation propagation in WD He shells have determined that the success of such propagation depends only upon the total WD mass and the available density of fuel. In other words, the success or failure of He shell detonation propagation in a single WD does not depend on the orientation of the collision, nor on the mass of the collision partner. Therefore the outcome of collision should only depend on the total mass and composition of each WD independently of one another.

\section{Discussion}\label{sec:sum}

\begin{figure}[]
\centering\includegraphics[width=\linewidth,clip=true]{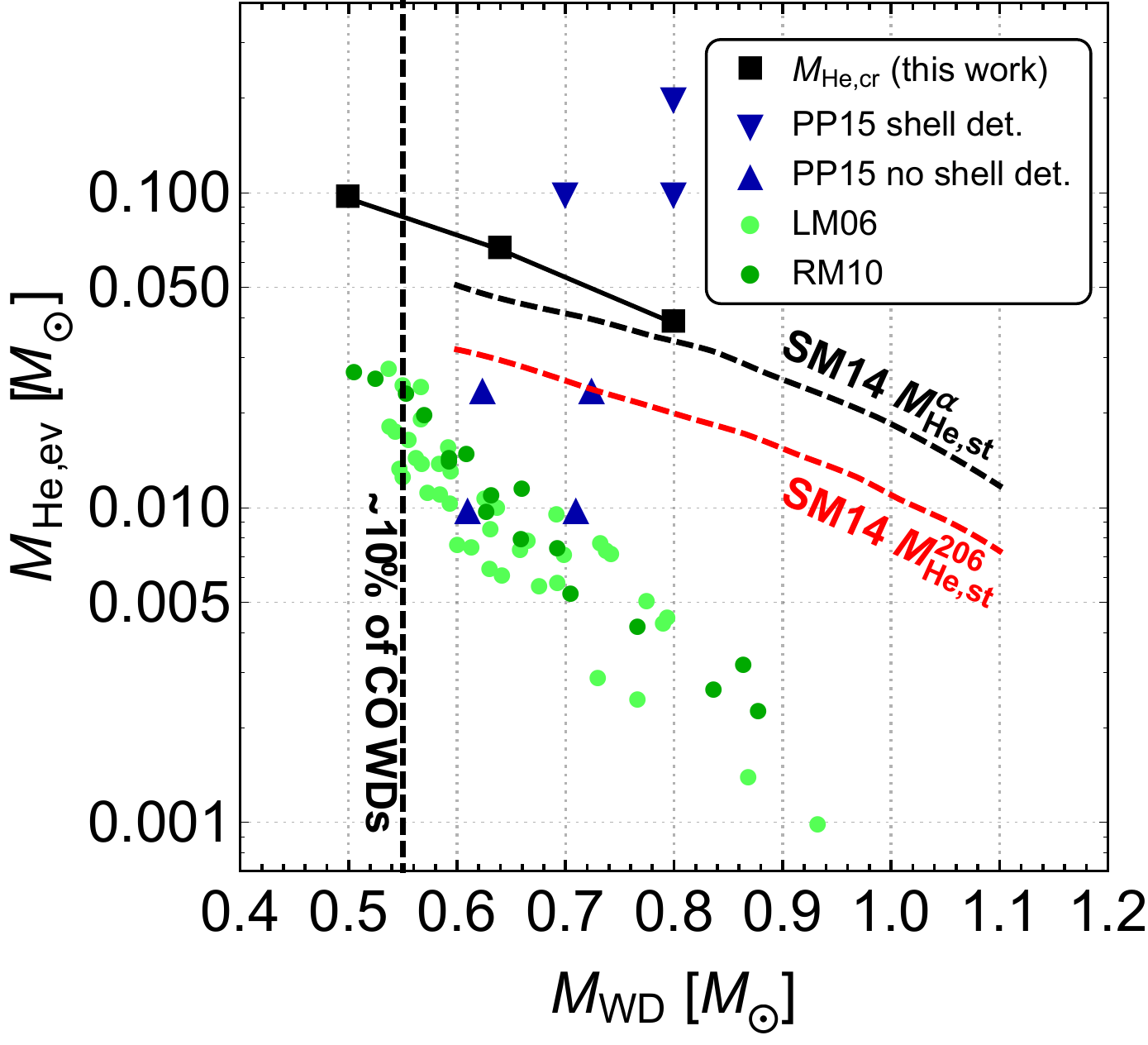}
\caption{The parameter space for WD helium shells. For perspective, WD He masses from evolutionary calculations $\mevol$ \citep{2006MNRAS.371..263L,2010ApJ...717..183R} are shown as green circles. Our calculations for $\mcross$ (Section \ref{sec:calib}) are displayed as black squares.  A selection drawn from the calculations of \cite{Papish:2015uy} is indicated by blue triangles, where WDs that supported He shell detonation and those that did not are distinguished by the orientation of the triangle. For comparison, we show the lines of minimal shell mass for supporting pure He detonations calculated by \cite{2014ApJ...797...46S}, both for an inaccurate 13 isotope $\alpha$-chain network ($M^{\alpha}_{\rm He,st}$; black, dashed) similar to the one utilized in this work, and a more sophisticated 206 isotope network ($M^{206}_{\rm He,st}$; red, dashed). The physics governing $M_{\rm He,st}$ and $\mcross$ is nearly the same, we therefore note that these quantities should be, and are, in good agreement (i.\ e.\  $\mcross\myapprox M^{\alpha}_{\rm He,st}$). If it can be assumed that $\mcross$ will align with $M^{206}_{\rm He,st}$ in the case that a larger network is employed, then the majority of WDs ($M_{\rm WD} \gtrsim 0.6\,M_{\odot}$) will have $\mevol < \mcross$ and thus are not significantly impacted by their intrinsic He components. The lowest mass CO WDs may have $\mevol$ in excess of $\mcross$. However, only $\myapprox10\%$ of CO WD have $\mwd\le0.55\,M_{\odot}$ \citep[vertical dashed black line;][]{2008AJ....135.1225H}. Therefore, even if low mass ($\myapprox 0.5 M_{\odot}$) WDs do have $\mevol > \mcross$, these collisions can only account for a small fraction of observed events.}
\label{fig:param}
\end{figure}

\begin{sidewaysfigure}[]
\centering\includegraphics[width= \linewidth,clip=true]{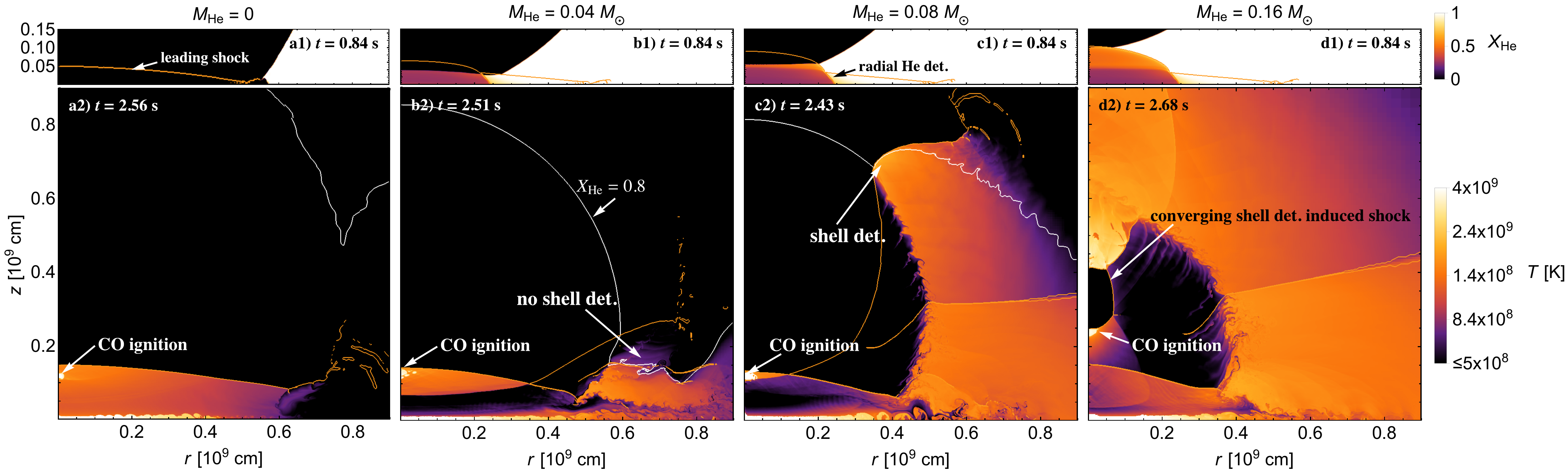}
\caption{Snapshots of the variety of collision behaviors as a function of He mass for $M_{\rm WD}=0.64 M_{\odot}$. In all panels the orange contours represent the location of shock fronts for matter at temperature $T>10^8$ K. Due to the intrinsic mirror symmetry of our 2D simulations, we only display the top half plane. Top panels show the He mass fraction $X_{\rm He}$. (a1) A leading shock wave propagates in the core of a pure CO WD ($M_{\rm He}=0$). (b1-d1) Same as panel a1, except now with non-zero $M_{\rm He}$. He burning in the contact zone accelerates the leading shock farther into the CO core compared to panel a1. A radially propagating He detonation forms. Bottom panels show logarithmic temperature maps, with white contours indicating $X_{\rm He}=0.8$. (a2) CO ignition in the pure CO WD. (b2) Same as panel a2, but the accelerated shock induces CO ignition at a slightly earlier time. The conditions of the unshocked He shell are not sufficient to faciliate the continued propagation of the He detonation, and it is quickly quenched. (c2) Same as panel b2. Additional He burning in the contact region results in further reduction of the CO ignition time. The He shell now extends sufficiently deep into the WD interior to support the propagation of the He detonation ignitied in the contact zone. The detonation sweeps around the shell, driving a curved shock into the WD interior. (d2) At larger $M_{\rm He}$, the CO is ignited behind the He shell induced shock, rather than the leading shock.}
\label{fig:shelldet}
\end{sidewaysfigure}

We conclude that it is unlikely that WD collisions will be significantly affected by their intrinsic He components. In section \ref{sec:0.64-0.64} we observed that the behavior of WD collisions may be appreciably altered from the results obtained in pure CO collisions provided that sufficient quantities of He exist in the outermost layers of the progenitors (Figure \ref{fig:shelldet}). We then empirically demonstrated that a helium content in excess of $\mcross$, the minimal mass for the He detonation to propagate into the unshocked shell, is a necessary condition for non-trivial modification of the ensuing CO ignition (section \ref{sec:calib}). Although we utilized a small reaction network which is known to be a poor approximation for He burning above $\sim$$10^9$ K, we have shown that our results are in good agreement with detailed studies concerning the impact of network size on He shell detonations. This agreement allowed us to infer the reductions in $\mcross$ we would obtain with a more sophisticated reaction network. Even with the enhancements provided by large nuclear networks, the minimal mass for supported He shell detonation $\msteady\myapprox\mcross$ is larger than the expected maximal He mass in WDs $\mevol$, except possibly for the lowest mass CO WDs ($M_{\rm WD}\myapprox 0.5\,M_{\odot}$) which are expected to contribute only a small fraction of collisions (Figure \ref{fig:param}).

One possible caveat is that real WDs are likely to have non-trivial compositional transition regions, wherein the helium layer is polluted with sizable quantities of carbon and oxygen, as well as smaller amounts of hydrogen and nitrogen \citep[][]{2010ApJ...717..183R}. \cite{2014ApJ...797...46S} showed that such pollutants can reduce  $\msteady$  by an additional $\sim$tens of percent from the pure He case when utilizing a large nuclear network. If this is indeed the case, it may be possible for the helium content of low mass WDs ($\myapprox 0.5 M_{\odot}$) to exceed $\mcross$. However, given that the composition profiles are complicated, and that their calculations remain somewhat uncertain, it is difficult to predict precisely how large the effect on $\mcross$ will be.

Finally, although the bulk properties of WD collisions are largely governed by the detonation of the CO core, He burning on the WD exterior can potentially produce observationally relevant isotopes \citep{2013ApJ...771...14H,2013ApJ...776...97M,Papish:2015uy}. The conditions characteristic of WD He shells are typically insufficient to produce $^{56}$Ni, and the mass within the shocked He shell is small, therefore nickel synthesis will not be noticeably changed unless the CO ignition time is reduced or delayed. However, intermediate mass elements such as $^{40}$Ca, $^{44}$Ti, and $^{48}$Cr can be produced in large quantities from the burnt He shell, but this again is reliant on the capacity for a given WD to support a He shell detonation. A serious effort to predict the nucleosynthesis of the He shell detonation would require a larger network than employed here. Taking into consideration that the largest uncertainties of this study, as well as those in other similar studies, stem from the use of abridged nuclear networks, we strongly urge the implementation of more sophisticated networks in future calculations concerning nuclear explosive astrophysics.

\acknowledgements
We thank B.\ Katz, K.\ Shen, J.\ Guillochon, and L.\ Althaus for useful discussions. C.~H. recognizes and appreciates support from the Department of Energy National Nuclear Security Administration Stewardship Science Graduate Fellowship under grant DE-NA0002135, and from Krell Institute. D.~K. gratefully acknowledges support from the Friends of the Institute for Advanced Study. FLASH was in part developed by the DOE NNSA-ASC OASCR Flash Center at the University of Chicago.

\bibliographystyle{apj}
\bibliography{cole}

\begin{thebibliography}{}
\expandafter\ifx\csname natexlab\endcsname\relax\def\natexlab#1{#1}\fi

\bibitem[{Althaus {et~al.}(2005)Althaus, Serenelli, Panei, C{\'o}rsico,
  Garc{\'\i}a-Berro, \& Sc{\'o}ccola}]{2005A&A...435..631A}
Althaus, L.~G., Serenelli, A.~M., Panei, J.~A., {et~al.} 2005, Astronomy and
  Astrophysics, 435, 631

\bibitem[{Aznar-Sigu{\'a}n {et~al.}(2013)Aznar-Sigu{\'a}n, Garc{\'\i}a-Berro,
  Lor{\'e}n-Aguilar, Jos{\'e}, \& Isern}]{2013MNRAS.434.2539A}
Aznar-Sigu{\'a}n, G., Garc{\'\i}a-Berro, E., Lor{\'e}n-Aguilar, P., Jos{\'e},
  J., \& Isern, J. 2013, Monthly Notices of the Royal Astronomical Society,
  434, 2539

\bibitem[{Benz {et~al.}(1989)Benz, Thielemann, \& Hills}]{1989ApJ...342..986B}
Benz, W., Thielemann, F.~K., \& Hills, J.~G. 1989, Astrophysical Journal, 342,
  986

\bibitem[{Bischoff-Kim {et~al.}(2014)Bischoff-Kim, {\O}stensen, Hermes, \&
  Provencal}]{2014ApJ...794...39B}
Bischoff-Kim, A., {\O}stensen, R.~H., Hermes, J.~J., \& Provencal, J.~L. 2014,
  The Astrophysical Journal, 794, 39

\bibitem[{Colgate \& McKee(1969)}]{1969ApJ...157..623C}
Colgate, S.~A., \& McKee, C. 1969, Astrophysical Journal, 157, 623

\bibitem[{Couch {et~al.}(2013)Couch, Graziani, \& Flocke}]{2013ApJ...778..181C}
Couch, S.~M., Graziani, C., \& Flocke, N. 2013, The Astrophysical Journal, 778,
  181

\bibitem[{Dong {et~al.}(2015)Dong, Katz, Kushnir, \&
  Prieto}]{2015MNRAS.454L..61D}
Dong, S., Katz, B., Kushnir, D., \& Prieto, J.~L. 2015, Monthly Notices of the
  Royal Astronomical Society: Letters, 454, L61

\bibitem[{Fryxell {et~al.}(2000)Fryxell, Olson, Ricker, Timmes, Zingale, Lamb,
  MacNeice, Rosner, Truran, \& Tufo}]{2000ApJS..131..273F}
Fryxell, B., Olson, K., Ricker, P., {et~al.} 2000, The Astrophysical Journal
  Supplement Series, 131, 273

\bibitem[{Hawley {et~al.}(2012)Hawley, Athanassiadou, \&
  Timmes}]{2012ApJ...759...39H}
Hawley, W.~P., Athanassiadou, T., \& Timmes, F.~X. 2012, The Astrophysical
  Journal, 759, 39

\bibitem[{Hillebrandt \& Niemeyer(2000)}]{2000ARA&A..38..191H}
Hillebrandt, W., \& Niemeyer, J.~C. 2000, Annual Review of Astronomy and
  Astrophysics, 38, 191

\bibitem[{Holberg {et~al.}(2008)Holberg, Sion, Oswalt, McCook, Foran, \&
  Subasavage}]{2008AJ....135.1225H}
Holberg, J.~B., Sion, E.~M., Oswalt, T., {et~al.} 2008, The Astronomical
  Journal, 135, 1225

\bibitem[{Holcomb {et~al.}(2013)Holcomb, Guillochon, De~Colle, \&
  Ramirez-Ruiz}]{2013ApJ...771...14H}
Holcomb, C., Guillochon, J., De~Colle, F., \& Ramirez-Ruiz, E. 2013, The
  Astrophysical Journal, 771, 14

\bibitem[{Kashyap {et~al.}(2015)Kashyap, Fisher, Garc{\'\i}a-Berro,
  Aznar-Sigu{\'a}n, Ji, \& Lor{\'e}n-Aguilar}]{2015ApJ...800L...7K}
Kashyap, R., Fisher, R., Garc{\'\i}a-Berro, E., {et~al.} 2015, The
  Astrophysical Journal Letters, 800, L7

\bibitem[{Katz \& Dong(2012)}]{2012arXiv1211.4584K}
Katz, B., \& Dong, S. 2012, arXiv, 4584

\bibitem[{Kozai(1962)}]{1962AJ.....67..591K}
Kozai, Y. 1962, Astronomical Journal, 67, 591

\bibitem[{Kushnir \& Katz(2014)}]{2014ApJ...785..124K}
Kushnir, D., \& Katz, B. 2014, The Astrophysical Journal, 785, 124

\bibitem[{Kushnir {et~al.}(2013)Kushnir, Katz, Dong, Livne, \&
  Fern{\'a}ndez}]{2013ApJ...778L..37K}
Kushnir, D., Katz, B., Dong, S., Livne, E., \& Fern{\'a}ndez, R. 2013, The
  Astrophysical Journal Letters, 778, L37

\bibitem[{Lawlor \& MacDonald(2006)}]{2006MNRAS.371..263L}
Lawlor, T.~M., \& MacDonald, J. 2006, Monthly Notices of the Royal Astronomical
  Society, 371, 263

\bibitem[{Lidov(1962)}]{1962P&SS....9..719L}
Lidov, M.~L. 1962, Planetary and Space Science, 9, 719

\bibitem[{Livne(1990)}]{1990ApJ...354L..53L}
Livne, E. 1990, Astrophysical Journal, 354, L53

\bibitem[{Lor{\'e}n-Aguilar {et~al.}(2009)Lor{\'e}n-Aguilar, Isern, \&
  Garc{\'\i}a-Berro}]{2009A&A...500.1193L}
Lor{\'e}n-Aguilar, P., Isern, J., \& Garc{\'\i}a-Berro, E. 2009, Astronomy and
  Astrophysics, 500, 1193

\bibitem[{Maoz {et~al.}(2014)Maoz, Mannucci, \& Nelemans}]{2014ARA&A..52..107M}
Maoz, D., Mannucci, F., \& Nelemans, G. 2014, Annual Review of Astronomy and
  Astrophysics, 52, 107

\bibitem[{Moll \& Woosley(2013)}]{2013ApJ...774..137M}
Moll, R., \& Woosley, S.~E. 2013, The Astrophysical Journal, 774, 137

\bibitem[{Moore {et~al.}(2013)Moore, Townsley, \&
  Bildsten}]{2013ApJ...776...97M}
Moore, K., Townsley, D.~M., \& Bildsten, L. 2013, The Astrophysical Journal,
  776, 97

\bibitem[{Papish \& Perets(2015)}]{Papish:2015uy}
Papish, O., \& Perets, H.~B. 2015, arXiv, 1502.03453v1

\bibitem[{Phillips(1993)}]{1993ApJ...413L.105P}
Phillips, M.~M. 1993, Astrophysical Journal, 413, L105

\bibitem[{Raskin {et~al.}(2010)Raskin, Scannapieco, Rockefeller, Fryer, Diehl,
  \& Timmes}]{2010ApJ...724..111R}
Raskin, C., Scannapieco, E., Rockefeller, G., {et~al.} 2010, The Astrophysical
  Journal, 724, 111

\bibitem[{Raskin {et~al.}(2009)Raskin, Timmes, Scannapieco, Diehl, \&
  Fryer}]{2009MNRAS.399L.156R}
Raskin, C., Timmes, F.~X., Scannapieco, E., Diehl, S., \& Fryer, C. 2009,
  Monthly Notices of the Royal Astronomical Society: Letters, 399, L156

\bibitem[{Renedo {et~al.}(2010)Renedo, Althaus, Miller~Bertolami, Romero,
  C{\'o}rsico, Rohrmann, \& Garc{\'\i}a-Berro}]{2010ApJ...717..183R}
Renedo, I., Althaus, L.~G., Miller~Bertolami, M.~M., {et~al.} 2010, The
  Astrophysical Journal, 717, 183

\bibitem[{Rosswog {et~al.}(2009)Rosswog, Kasen, Guillochon, \&
  Ramirez-Ruiz}]{2009ApJ...705L.128R}
Rosswog, S., Kasen, D., Guillochon, J., \& Ramirez-Ruiz, E. 2009, The
  Astrophysical Journal Letters, 705, L128

\bibitem[{Shen \& Bildsten(2009)}]{2009ApJ...699.1365S}
Shen, K.~J., \& Bildsten, L. 2009, The Astrophysical Journal, 699, 1365

\bibitem[{Shen \& Bildsten(2014)}]{2014ApJ...785...61S}
---. 2014, The Astrophysical Journal, 785, 61

\bibitem[{Shen \& Moore(2014)}]{2014ApJ...797...46S}
Shen, K.~J., \& Moore, K. 2014, The Astrophysical Journal, 797, 46

\bibitem[{Thompson(2011)}]{2011ApJ...741...82T}
Thompson, T.~A. 2011, The Astrophysical Journal, 741, 82

\bibitem[{Timmes \& Swesty(2000)}]{2000ApJS..126..501T}
Timmes, F.~X., \& Swesty, F.~D. 2000, The Astrophysical Journal Supplement
  Series, 126, 501

\bibitem[{Townsley {et~al.}(2012)Townsley, Moore, \&
  Bildsten}]{2012ApJ...755....4T}
Townsley, D.~M., Moore, K., \& Bildsten, L. 2012, The Astrophysical Journal,
  755, 4

\bibitem[{Weaver {et~al.}(1978)Weaver, Zimmerman, \&
  Woosley}]{1978ApJ...225.1021W}
Weaver, T.~A., Zimmerman, G.~B., \& Woosley, S.~E. 1978, Astrophysical Journal,
  225, 1021

\bibitem[{Weinberg {et~al.}(2006)Weinberg, Bildsten, \&
  Schatz}]{2006ApJ...639.1018W}
Weinberg, N.~N., Bildsten, L., \& Schatz, H. 2006, The Astrophysical Journal,
  639, 1018

\bibitem[{Woosley \& Kasen(2011)}]{2011ApJ...734...38W}
Woosley, S.~E., \& Kasen, D. 2011, The Astrophysical Journal, 734, 38

\end{thebibliography}


\begin{table*}[t]
\centering
\begin{tabular}{ccccc}
\hline
 $M_1=M_2$ & $\mhe$ &  detonation in the &  detonation in the  & CO ignition time\\
 $[M_{\odot}]$ & $[10^{-3}\,M_\odot]$ & shocked helium layer & un-shocked helium layer  &[$\textrm{s}$]\\
\hline
\hline
0.5 & 0 & -- & -- & 4.54 \\ 
0.5 & 10 & x& -- & 4.45 \\ 
0.5 & 15 & \checkmark & x& 4.45 \\ 
0.5 & 20 & \checkmark & x& 4.46  \\ 
0.5 & 50 & \checkmark & x& 4.54 \\ 
0.5 & 80 & \checkmark & x & 4.40 \\ 
0.5 & 90 & \checkmark & x & 4.65 \\ 
0.5 & 91 & \checkmark & x & 4.82  \\ 
0.5 & 92 & \checkmark & x & 4.87 \\ 
0.5 & 93 & \checkmark & x & 4.91 \\ 
0.5 & 94 & \checkmark & \checkmark & 4.45 (dd) \\ 
0.5 & 95 & \checkmark & \checkmark & 4.35 (dd) \\ 
0.5 & 96 & \checkmark & \checkmark & 4.35 (dd) \\ 
0.5 & 100 & \checkmark & \checkmark & 4.28 (dd) \\ 
0.5 & 120 & \checkmark & \checkmark & 4.00 (dd) \\ 
0.5 & 140 & \checkmark & \checkmark & 3.83 (dd) \\ 
0.5 & 160 & \checkmark & \checkmark & 3.68 (dd) \\
0.5 & 180 & \checkmark & \checkmark & 3.55 (dd) \\
0.5 & 200 & \checkmark & \checkmark & 3.43 (dd) \\
\hline    
\end{tabular}
\caption[]{Simulation outcomes for the $0.5-0.5$ case. The symbol ${\rm (dd)}$ indicates a double detonation-like CO ignition (see text for details).}
\label{table:toy}
\end{table*}

\begin{table*}[t]
\centering
\scalebox{0.75}{
\begin{tabular}{ccccc}
\hline
 $M_1=M_2$ & $\mhe$ &  detonation in the &  detonation in the  & CO ignition time\\
 $[M_{\odot}]$ & $[10^{-3}\,M_\odot]$ & shocked helium layer & un-shocked helium layer  &[$\textrm{s}$]\\
\hline
\hline
0.64 & 0 & -- & -- & 2.56 \\
0.64 & 5 & x & -- & 2.49  \\
0.64 & 6 & x & -- & 2.47  \\
0.64 & 7 & x & -- & 2.47  \\
0.64 & 8 & \checkmark & x & 2.44  \\
0.64 & 10 & \checkmark & x & 2.50  \\
0.64 & 20 & \checkmark & x & 2.55  \\
0.64 & 40 & \checkmark & x & 2.51\\
0.64 & 60 & \checkmark & x & 2.40  \\
0.64 & 65 & \checkmark & x & 2.39 \\
0.64 & 66 & \checkmark & x & 2.40  \\
0.64 & 67 & \checkmark & \checkmark & 2.40  \\
0.64 & 68 & \checkmark & \checkmark & 2.41  \\
0.64 & 69 & \checkmark & \checkmark & 2.41  \\
0.64 & 70 & \checkmark & \checkmark & 2.41  \\
0.64 & 80 & \checkmark & \checkmark & 2.43 \\
0.64 & 90 & \checkmark & \checkmark & 2.44 \\
0.64 & 100 & \checkmark & \checkmark & 2.49  \\
0.64 & 110 & \checkmark & \checkmark & 2.56  \\
0.64 & 120 & \checkmark & \checkmark & 2.63  \\
0.64 & 130 & \checkmark & \checkmark & 2.70  \\
0.64 & 140 & \checkmark & \checkmark & 2.76 (dd+leading) \\
0.64 & 150 & \checkmark & \checkmark & 2.72 (dd) \\
0.64 & 160 & \checkmark & \checkmark & 2.68 (dd) \\
0.64 & 180 & \checkmark & \checkmark & 2.60 (dd) \\
0.64 & 200 & \checkmark & \checkmark & 2.50 (dd) \\
\hline
\end{tabular}
}
\caption[]{Simulation outcomes for $0.5-0.5$ and $0.64-0.64$ cases. The symbol ${\rm (dd)}$ indicates a double detonation-like CO ignition, and ${\rm (dd + normal)}$ indicates that the normal leading shock ignition and double detonation-like CO ignition (see text for details) occured simultaneously.}
\label{table:toy2}
\end{table*}

  \begin{table*}[t]
\centering\begin{tabular}{ccccc}
\hline
 $M_1=M_2$ & $\mhe$ &  detonation in the &  detonation in the  & CO ignition time\\
 $[M_{\odot}]$ & $[10^{-3}\,M_\odot]$ & shocked helium layer & un-shocked helium layer  &[$\textrm{s}$]\\
\hline
\hline
0.8 & 0 & -- & -- & 1.36\\
0.8 & 5 & x & -- & 1.34  \\
0.8 & 10 & \checkmark & x & 1.40  \\
0.8 & 20 & \checkmark & x & 1.49  \\
0.8 & 30 & \checkmark & x & 1.43 \\
0.8 & 38 & \checkmark & x & 1.42  \\
0.8 & 39 & \checkmark & x & 1.41 \\
0.8 & 40 & \checkmark & \checkmark & 1.41  \\
0.8 & 41 & \checkmark & \checkmark & 1.41 \\
0.8 & 42 & \checkmark & \checkmark & 1.41 \\
0.8 & 50 & \checkmark & \checkmark & 1.41  \\
0.8 & 60 & \checkmark & \checkmark & 1.38  \\
0.8 & 80 & \checkmark & \checkmark & 1.44 \\
0.8 & 100 & \checkmark & \checkmark & 1.56 \\
0.8 & 120 & \checkmark & \checkmark & 1.62  \\
0.8 & 140 & \checkmark & \checkmark & 1.71 \\
0.8 & 160 & \checkmark & \checkmark & 1.80 \\
0.8 & 180 & \checkmark & \checkmark & 0.52 (di) \\
0.8 & 200 & \checkmark & \checkmark & 0.53 (di) \\
\hline
\end{tabular}
\caption[]{Simulation outcomes for the $0.8-0.8$ case. The symbol ${\rm (di)}$ indicates a direct ignition (see text for details).}
\label{table:toy3}
\end{table*}

\end{document}